\documentclass[11 pt,a4paper]{article}
\usepackage[utf8]{inputenc}
\usepackage{amsmath}
\usepackage{amsfonts}
\usepackage{amssymb}
\usepackage{pgfplots}
\usepackage{tikz}
\usepgfplotslibrary{fillbetween}
\usetikzlibrary{patterns}
\usepackage{indentfirst}
 
\author{Dr. Jacques Balayla MD, MPH\footnote{To whom correspondence should be addressed: Dr. Jacques Balayla MD, MPH. Quilligan Scholar. e-mail: jacques.balayla@mail.mcgill.ca. Department of Obstetrics and Gynaecology. McGill University, Montreal, Quebec, Canada}}

\title{The Individual Impact Index ($i^3$) Statistic: \\A Novel Article-Level Citation Metric}
\date{}

\begin{document}
\maketitle  

\begin{abstract}

	Citation metrics are analytic measures used to evaluate the usage, impact and dissemination of scientific research. Traditionally, citation metrics have been independently measured at each level of the publication pyramid, namely: at the article-level, at the author-level, and at the journal-level. The most commonly used metrics have been focused on journal-level measurements, such as the impact factor (IF) and the Eigenfactor, as well as on researcher-level metrics like the Hirsch index (h-index) and i10-index. On the other hand, reliable article-level metrics are less widespread, and are often reserved to non-standardized characteristics of individual articles, such as views, citations, downloads, and mentions in social and news media. These characteristics are known as “altmetrics”. However, when the number of views and citations are similar between two articles, no discriminating measure currently exists with which to assess and compare each article’s individual impact. Given the modern exponentially-growing scientific literature,  scientists and readers of science need reliable, objective methods for managing, measuring and comparing research outputs and publications. To this end, I hereby describe and propose a new standardized article-level metric henceforth known as the “Individual Impact Index ($i^3$) Statistic”. The $i^3$ is a weighted algorithm that takes advantage of the peer-review process, and considers a number of characteristics of individual scientific publications in order to yield a standardized and readily comparable measure of impact and dissemination. The strengths, limitations and potential uses of this novel metric are also discussed.
\end{abstract} 

\newpage

\section{Introduction} In scholarly and scientific publishing, citation metrics are used to evaluate the usage and impact of scientific research \cite{neylon2009article}. Traditional citation metrics have usually been focused on journal-level measurements such as the impact factor (IF) and the Eigenfactor, as well as on researcher-level metrics like the Hirsch index (h-index) \cite{bergstrom2008eigenfactor, bornmann2005does, garfield2006history}. On the other hand, standardized article-level metrics are less widespread, and are often reserved to non-standardized characteristics of individual articles, such as views, citations, downloads, and mentions in social and news media \cite{thelwall2013altmetrics}. These are commonly referred to as “altmetrics”, and provide a record of attention, a measure of dissemination and an indicator of influence and impact of a given article \cite{thelwall2013altmetrics}. However, there are important limitations to such article-level metrics since they do not distinguish between self-views and views by others, downloads by scientists or by laypersons, and self-disseminations through social media or press releases \cite{bornmann2014altmetrics}. 
Additionally, when the number of views and citations are similar between two articles, no discriminating measure currently exists with which to assess and compare each article's individual impact. Yet, two publications with a similar number of citations can vary immensely in terms of quality and reach. Indeed, a comprehensive tool designed to appraise the individual impact of scholarly works is necessary to aid in such assessment. Given the aforementioned considerations, I hereby describe and propose a new article-level metric henceforth known as the “Individual Impact Index ($i^3$) Statistic”: a weighted algorithm that takes into account the scientific source and domain of the publication, the number of citations, as well as the provenance of those citations in order to yield a standardized and readily comparable measure of impact and dissemination for scholarly publications. 

\subsection{What is understood by \textit{Impact} in Science?}
The proper meaning of ‘impact’ in Science is not conveyed by a single definition. Rather, how one defines impact in Science may vary depending on the circumstance. A scholarly work that describes a new discovery or potential novel solution to a longstanding problem may change the way in which a scientific or technological process is undertaken. Certainly, that is impactful in the proper sense. However, in publication metrics, impact is not traditionally measured by the effect of a particular research, but rather, impact can more readily be defined as a measure of dissemination and reach of the scientific information in question \cite{bollen2009principal}. Although works of major impact are invariably well disseminated, characterizing and standardizing that reach is a difficult process to undertake, as no reliable method or tool currently exists. The $i^3$ statistic seeks to fill that void.

\subsection{What are Citation Metrics?} How do we then assess quality in scientific research and scholarly publications? The only true way to assess quality would be to subject and compare individual publications to a given standard, assigning an arbitrary measure of quality to a number of characteristics shared with the standard by the publication in question, such as presentation, methodology, and validity of conclusions \cite{neylon2009article}. However, imagine that as an avid reader of Science, you seek to \textit{learn} something new. How can a novice assess the quality of a publication on a subject he/she knows nothing about? Even if an individual mastered the topic at hand, establishing true quality is cumbersome, as it would imply having to read, analyse, and re-test every conclusion in the publication in question. Instead, we establish subjective quality with the help of citation metrics \cite{harzing2010citation}. 
Citation metrics are specific measurements of activity that quantify the usage and dissemination of scholarly works \cite{harzing2010citation}. Greater usage implies that other scientists and individuals with expertise have accessed, shared and cited the publication in question. Because greater usage implies greater acceptance by fellow scientists, usage and dissemination are commonly used as a proxy for quality in Science. 

\subsection{Levels of Citation Metrics} Citation metrics can be studied at different levels in the publication pathway: the journal, the author and the article \cite{neylon2009article, garfield2006history, bornmann2005does}. Essentially, the same question is asked at each level: is the journal/author/paper good/impactful? Journal-level metrics reveal the influence of a journal in communicating the most relevant research. Most scientists will point to Thomson Reuters's Journal of Citation Reports (JCR) \textit{Impact Factor} (IF) as an external and objective measure for ranking the impact of specific journals \cite{garfield2006history}. The impact factor is a measure reflecting the yearly average number of citations to recent articles published in a given journal. Journals with higher impact factors are often deemed to be more important and reputable than those with lower ones. Though their use is less widespread, other journal-level metrics include the Eigenfactor and the 5-year Impact Factor. Author-level metrics are less widespread than journal-level metrics. They serve a more indirect purpose than journal- or article-level metrics. These metrics allow for comparisons between researchers on productivity and impact. While these may aid in funding of grants, distribution of resources, and hiring decisions, they are traditionally of little use to other scientists. The \textit{Hirsch} Index (h-index) is the main author-level metric used today. It is a measure reflecting the maximum number of articles in an author's repertoire having at least that same number of citations. On the other hand, as mentioned previously, scientifically rigorous article-level metrics are less widespread, and are often reserved to non-standardized characteristics of individual articles. Although many attempts at developing new article-level metrics have been undertaken, the number of citations remains at the helm of the majority of these.

\subsection{Journal of Citation Reports (JCR)}
The Journal Citation Reports (JCR) is an annual publication by Clarivate Analytics (previously the Intellectual Property and Science business of Thomson Reuters), which provides systematic and objective means to evaluate the world's leading scientific and scholarly journals \cite{garfield2006history}. The JCR is an authoritative resource for impact factors in academic journals in the sciences (SCIE) and social sciences (SSCI) disciplines, publishing annual information on impact factors and other journal-level metrics, such as the immediacy index \cite{tomer1986statistical}. Other information provided by the JCR includes basic bibliographic information of the journals such as the publisher, title abbreviation, language, and ISSN identification. Furthermore, JCR categorizes journals into 171 categories in the sciences and 54 in the social sciences. Simply put, these subject categories refer to the different disciplines into which academic journals are classified \cite{dorta2013comparing}. The impact factor information allows therefore for each journal to be ranked within its own subject category.

\section{The $i^3$ Statistic\protect\footnote{The $i^3$ Statistic is provisionally patent protected through the United States Patent and Trademark Office under the serial: USPTO 62/506,119}}
The $i^3$ statistic is hereby proposed as the weighted measure of the impact of an individual article. Currently, none of the established article-level metrics provide a distinction between the number of citations and the impact of an article. Instead they are often used as a proxy for one another. The $i^3$ statistic is a measurement that for any given publication, considers: the JCR category of the journal where the article is published, the impact factor of the publishing journal, as well as the number of citations and the provenance of those citations, assigning higher value or “impact” to citations in journals with higher \textit{impact factors} than lower ones. If we use the number of citations as an objective measure of impact then one could argue that a paper published in a remote, unknown journal with \textit{x} citations has more impact than a paper in a prestigious journal with \textit{x-b} citations. Intuitively, we know this to not systematically be the case. When considering the impact of a scholarly work, citation number textit{and} provenance should be equally important to rank publications. The statistic's algorithm accounting for the aforementioned considerations is the following:
 
\begin{equation}
i^3 = G(f(x)) = 1-e^{-\beta{f(x)}}
\end{equation}
where
\begin{equation}
f(x) = \psi_a + \sum_{i=0}^x\ \eta_x\psi_x = \psi_a + \eta_1\psi_1 + \eta_2\psi_2 +... + \eta_x\psi_x
\end{equation}
and 
\begin{equation}
\beta = \frac{1}{3\pi \phi} 
\end{equation} 

\
\

Equation (1) depicts the main algorithm to calculate the $i^3$ statistic index. \textit{e} refers to Euler's number, an irrational and transcendental constant playing a crucial role in mathematics and number theory, which can de defined as follows:

$$e = \sum_{n=0}^{\infty}  \dfrac{1}{n!} = \dfrac{1}{0!} + \dfrac{1}{1!} + \dfrac{1}{2!} +...+ \dfrac{1}{n!}+ \dfrac{1}{n+1!} + ... $$ 

\

$\beta$ refers to the \textit{Balayla} coefficient, which is described below. Equation (2) is a linear equation, which takes into account the impact factor of the publishing journal ($\psi_a$) as well as the sum product of citation number and provenance, counting the individual number of citations $\eta$ and the impact factor of the journals where they are found ($\psi_x$). As stated before, Equation (3) depicts the \textit{Balayla} coefficient, a JCR category-specific, unitless coefficient, which is inversely proportional to \textit{$\phi$}, the number of journal titles in a given JCR category.

\subsection{Properties of the $i^3$ Statistic}

The equation for the $i^3$ statistic is complement to the exponential decay function, with an asymptote at y = 1, the maximum theoretical $i^3$ value a scientific paper can have. Given the above equation (1), it follows that: 

$$\lim_{f(x)\to 0} G(f(x)) = 0$$
and
$$\lim_{f(x)\to\infty} G(f(x)) = 1$$
\
\

As such, the $i^3$ yields a value between 0 and 1. The $i^3$ = 0 when no citations have taken place, and the $i^3$ = 1 when, theoretically, a single article has infinite citations. Articles with an $i^3$ value closer to 1 are deemed to have more impact than those closer to 0. By standardizing the citation information the $i^3$ can be used as a ranking algorithm and as a tool to optimize literature searches. Following is the graphic representation of the $i^3$ function, for an average $\beta$ coefficient of 0.00115, which corresponds to a JCR category with $\phi$ = 92 titles.

\subsection{Graphic representation of the $i^3$ Statistic\protect\footnote{assumes an average $\beta$ coefficient of 0.0011.}}

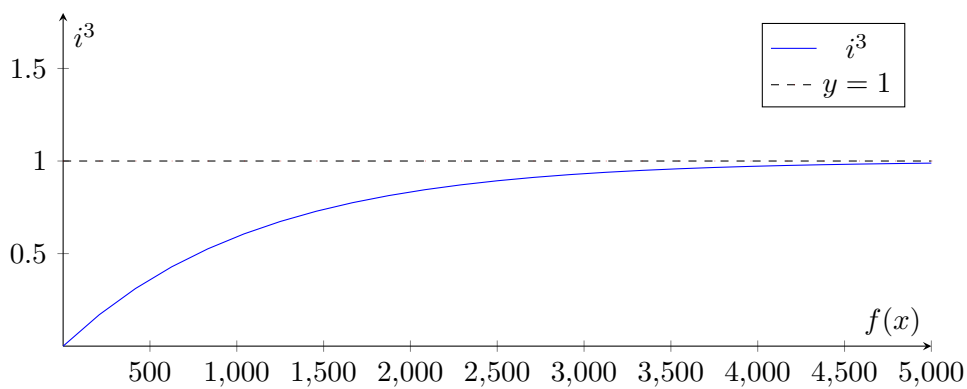
\begin{center}

\begin{tikzpicture}
 
	\begin{axis}[
    axis lines = center,
    xlabel = $f(x)$,
	ylabel = {$i^3$},    
     ymin=0, ymax=1.8,
    legend pos = north east,
     ymajorgrids=false,
    grid style=dashed,
    width=13cm,
    height=6cm,
     ]
	\addplot [
	domain= 0:5000,
	color= blue,
	]
	{1-2.1778^(-0.00115*x)};
\addplot+ [
dashed,
domain= 0:5000,	
color = black,
mark size = 0pt
 ]
	{1};
	\addlegendentry{$i^3$}
	\addlegendentry{$y=1$}
	\end{axis}

\end{tikzpicture}
\end{center}

\begin{center} Note the slow rise of the $i^3$ statistic as a function of \textit{f(x)} in blue, and the asymptote at y=1 in the black dashed line.
\end{center}

\subsection{The Balayla ($\beta$) Coefficient}

The \textit{Balayla} coefficient, an eponym named after the author, constitutes a positive rational number, which assigns a value as an inverse function of the number of journal titles \textit{$\phi$} in a given JCR category. By definition, \textit{$\phi$} is a natural integer greater or equal to 1, as the journal where the article in question is found yields at a minimum, a value of \textit{$\phi$} = 1. The purpose of the $\beta$ coefficient is to compensate $i^3$ scores in JCR categories where the number of titles is lower, thereby lowering opportunity to publish in domain-specific titles. Though articles can be published in general journals that fit domains other than their own, preliminary analysis shows that the vast majority of research articles are published in journals within their own domain categories. Given the above equation (3), it follows that:  

$${\lim_{\phi\to 1} \beta \simeq 0.1}$$ 
and 
$$\lim_{\phi\to\infty} \beta = 0$$

\begin{center}
The $\beta$ coefficient therefore takes on values between 0 and $\approx0.1$
\end{center}

\subsection{How was the Balayla Coefficient determined?}

The Balayla coefficient was determined empirically using real data from JCR's 2015 report. A preliminary search determined that in the citation distribution curve of all publications, 1000 citations falls on average at the 90th-95th percentile. A corresponding $i^3$ value of 0.90 should evoke a similar percentile in the $i^3$ score distribution. I have calculated an mean value of $\psi$ = 2.00 for the impact factor of a average journal, leading to f(x) $\approx 2,000$. Similar calculations reveals the average number of journals per  JCR category to be $\phi$ = 79.4. Isolating $\beta$ and inputting the above values we obtain:

$$i^3 = G(f(x)) = 1-e^{-\beta{f(x)}}$$
$$0.90 = G(f(x)) = 1-e^{-{2000}\beta}$$
$$0.1 = e^{-2000\beta}$$
$$ln(0.1) = ln(e^{-2000\beta})$$
$$\beta \approx 0.00115$$

\
\\
Let $\lambda$ be the coefficient relating $\beta$ to $\phi$. It therefore follows that:
\\

\begin{equation}
\boxed{\lambda = \beta\phi  \leftrightarrow  \beta=\frac{\lambda}{\phi}}
\end{equation}
\\

The inverse relationship between an average $\beta$ coefficient = 0.00115 and an average number of titles per category in the JCR $\phi$ = 79.4 leads to the following determination:
\

$$0.00115 = \dfrac{1}{79.4x} \Rightarrow \lambda\approx\dfrac{1}{3\pi}\Leftrightarrow\boxed{\beta=\frac{1}{3\pi\phi}}$$
\\
 
\subsection{Density distribution of the Balayla Coefficient}

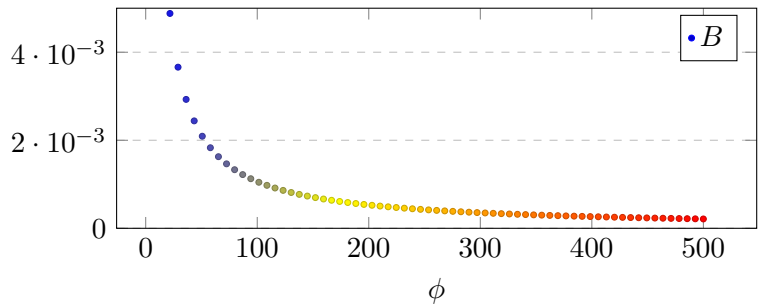
\begin{center}

\begin{tikzpicture}
	\begin{axis} 
	[
    axis lines = box,
    xlabel = $\phi$,
    ylabel = {},
     ymin=0, ymax=0.005,
     domain= 0:500,
     ymajorgrids=true,
     grid style=dashed,
     mark size = 1.1 pt,
     width=10cm,
    height=4.5cm,
    legend pos = north east
     ]
     
	\addplot+[scatter,only marks,
		 samples=70,scatter src=x]
		{1/(3*pi*x)};
	\addlegendentry{$B$}
	\pgfplotsset{scaled y ticks=false}
	\end{axis}
\end{tikzpicture}

\end{center}

\subsection{Dynamics of the $i^3$ Statistic}
The $i^3$ serves multiple purposes. One of these, is allowing the tracking of an individual paper's impact over time through fluctuations in the $i^3$ score. We define the following notations in this regard:
\\
\begin{center}
$i^3_x \rightarrow i^3_1, i^3_5, i^3_{10}$
\end{center}

where,
\\
\\
$i^3_1$ = An article's $i^3$ score one year after its original date of publication.
\\
\\
$i^3_5$ = An article's $i^3$ score five years after its original date of publication.
\\
\\
$i^3_{10}$ = An article's $i^3$ score ten years after its original date of publication.
\\
\\

The $i^3_{CR(t)}$ stands for the $i^3$ citation ratio at time \textit{t}. We can use the $i^3_t$ formulation above, where t = year since original publication off and divide it by the total $i^3$ score amassed by the publication throughout its history up until the point of interest. Of importance, this ratio uses information about the number and provenance of citations and journal impact factors both at time \textit{t} for the numerator and at the present time for the denominator. We can use this information to create distribution curves of citations, and thus determine whether the impact of a scholarly work is evenly distributed throughout its history or whether its concentrated at any one individual or multiple points.

\begin{equation}
i^3_{CR(t)} = \dfrac{i^3_t}{i^3}
\end{equation}
\\

However, because the $i^3$ is not a linear function a more precise measure of the ratio $i^3_{CR(t)}$ and citation distribution is obtained by integrating each of the $i^3$ functions, namely  $i^3_t$ and $i^3$, and dividing them to obtain the $i^3_{CR(t)}$.
\\
\\
\begin{equation}
i^3_{CR(t)} = \dfrac{\int _{ 0 }^{ t } { { { i }_{t}^{ 3 } } }  dt}{\int _{ 0 }^{ f(x) }{ { { i }^{ 3 } } } df(x)}
\end{equation}
\\
\\
\begin{equation}
i^3_{CR(t)}=\frac{\frac{1}{\beta}e^{{-\beta}{t}} + t + C}{\frac{1}{\beta}e^{-{\beta}f(x)} +f(x) + C}
\end{equation}

The algorithm for the $i^3$ remains the same. The f(x) equation is replaced by the variable \textit{t} in the numerator to denote the period of interest up to which the $i^3_{CR(t)}$ is being calculated.

\begin{equation}
t = \psi_a + \sum_{i=0}^t\ \eta_t\psi_t = \psi_a + \eta_1\psi_1 + \eta_2\psi_2 +... + \eta_t\psi_t
\end{equation}
\\
By definition, regarding equations (2) and (7):
\\
\begin{equation}
\sum_{i=0}^t\eta_t\psi_t \leq \sum_{i=0}^x\eta_x\psi_x
\end{equation}
\

Considering the solved integral in equation (6) we deduce that as the number of citations increases to infinity, the $i^3_{CR(t)}$ approaches the simple ratio of the sum-product of the number and provenance of citations, as stipulated in equation (2): 

\begin{equation}
\lim _{ x,t\rightarrow \infty  }{ \frac { { \int _{ 0 }^{ t }{ { I }_{ t }^{ 3 } } dt } }{ \int _{ 0 }^{ f(x) }{ { I }^{ 3 } } df(x) }  } =\frac { t }{ f(x) } =\frac { { f(x) }_{ t } }{ f(x) }
\end{equation}

\subsection{Circling around the asymptote}

As shown in the graphic representation of the $i^3$ statistic (Section 2.2), the curve reaches an asymptote at y=1 when f(x) approaches $\approx$ 3,500. The latter is of course dependent on the $\beta$ coefficient, and therefore, of the subject category of the article in question. Given the asymptote of the $i^3$ function, the derivative approaches 0 as f(x) goes to infinity. 

\begin{equation}
G'f(x) = i^3\dfrac{d}{dx} = 1 - e^{-\beta{f(x)}}\dfrac{d}{dx} 
\end{equation}

\begin{equation}
G'f(x) = \beta{e^{-\beta{f(x)}}}
\end{equation}

\begin{equation}
\lim_{f(x)\to\infty } G'(f(x)) = 0
\end{equation}
\

In other words, the f(x) of a publication increases at a slower rate beyond a certain point. Though in practical terms this means the impact of the publication in question is very large, it makes comparison with other like-publications more cumbersome. To get around this limitation, we can integrate the functions again to determine whether the small difference in $i^3$ values corresponds to actual large discrepancies in impact.

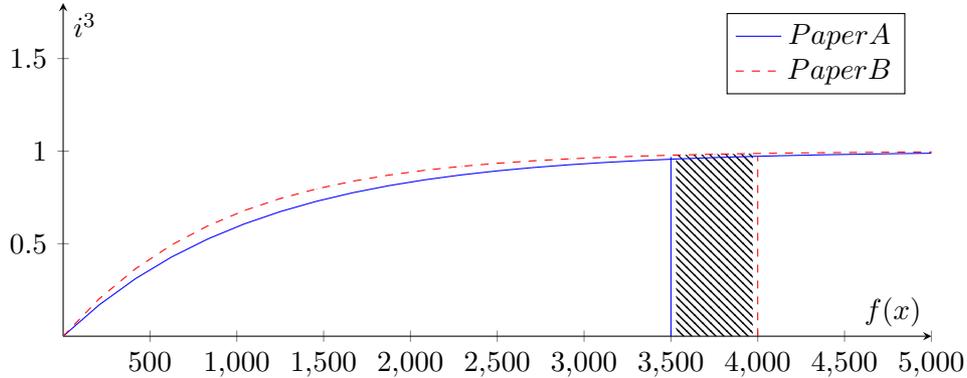
\begin{center}

\begin{tikzpicture}
 
	\begin{axis}[
    axis lines = center,
    xlabel = $f(x)$,
	ylabel = {$i^3$},    
     ymin=0, ymax=1.8,
    legend pos = north east,
     ymajorgrids=false,
    grid style=dashed,
    width=13cm,
    height=6cm,
    fill = blue,
     ]
	\addplot [
	domain= 0:5000,
	color= blue,
	name path = A
	]
	{1-2.1778^(-0.00115*x)};
\addplot+ [
dashed,
domain= 0:5000,	
color = red,
mark size = 0pt,
name path = B
 ]
	{1-2.1778^(-0.00140*x)};
	
\addplot +[mark=none, color=blue] coordinates {(3500, 0) (3500, 0.9687)};

\addplot +[mark=none, color=red, dashed] coordinates {(4000, 0) (4000, 0.9687)};
\path[name path=xaxis] (\pgfkeysvalueof{/pgfplots/xmin}, 0) -- (\pgfkeysvalueof{/pgfplots/xmax},0);

\addplot[red, pattern=north west lines] fill between[of=B and xaxis, soft clip={domain=3530:3970}];

\addplot[blue, pattern=north west lines] fill between[of=A and xaxis, soft clip={domain=3530:3970}];
      
\addlegendentry{$Paper A$}
\addlegendentry{$Paper B$}
\end{axis} 

\end{tikzpicture}
\end{center}

\begin{center}
\begin{small}

Note how for an f(x) = 3500 (Paper A) and an f(x) = 4000 (Paper B), the  $i^3$ values are similar. Integrating the $i^3$ functions and calculating the area under the curve (AUC) facilitates ranking and comparison. The dashed area in grey demonstrates excess area of paper B, indicating its impact is greater.

\end{small}
\end{center}

As noted, the area under the curve (AUC) measurements at the extreme of the $i^3$ function can assist in the comparison of highly cited works. This calculation serves as an adjunct measure of the dynamic history of the publication, in that the AUC increases with increasing f(x), independently of the function's ($i^3$) asymptote. 
\
\subsection{Rationale behind the $i^3$ Statistic}

The main driving force behind the development of the $i^3$ algorithm was the need for a optimized method to navigate the exponentially-growing scientific literature. Simply put, faced with several articles on a similar topic, a simple measure such as the $i^3$, a index number between 0 and 1, can allow for the ranking of said publications based on their perceived impact, indicating which is likely to be of higher quality. Because the $i^3$ is based on the number and provenance of citations in other scientific sources that have likely undergone peer review, it can effectively amass the validation and approval of a whole community of experts who have read the publication in question and considered it worth of a citation in their own work. Indeed, relative to non-scientific publications, which are not regulated and where no standardized evaluation of content takes place, the $i^3$ statistic takes advantage of peer-review and peer-expertise to evaluate and validate research findings, which are then disseminated through citations. If journals with a higher impact factor are deemed more reputable than those with lower impact factors, then it can be argued that the articles cited in those journals ought to be considered more reputable than those which aren't cited. Similarly, if journals with a higher impact factor are deemed more reputable, then the readership of those journals ought to be greater, and so is the exposure of articles cited in those journals. Finally, if journals with a higher impact factor are presumably more difficult to publish in, then the articles cited in those journals are likely subject to a higher standard, thus better reflecting their true worth. In one sentence, the premise of the $i^3$ is that an individual article's impact is a \textit{composite} measure of its own original publication and the characteristics of its citation history.

\section{Uses, Strengths and Limitations of the $i^3$}

The potential uses of the $i^3$ statistic are numerous. First, it can stratify scientific work by a rigorous measure of dissemination and perceived quality. Secondly, it can be used as a screening litmus test to assess a scientist's work and as a tool to rank a scientist's portfolio. It is conceivable that by allowing for the ranking of individual publications, the $i^3$ may aid in optimizing literature searches and reading tasks, and more generally, establishing a hierarchy to attribute awards, funding and grant money when funds are limited. Similarly, since the $i^3$ will be category-specific, it can potentially be used to develop reference values of impact percentiles by discipline, allowing scientists to set reasonable objectives for the expected impact of their work. Furthermore, journals can use the $i^3$ to estimate a distribution of the citation pattern of the articles they publish. Finally, the $i^3$ may be used to counter the "Matthew Effect", a primarily sociological phenomenon whereby in certain scenarios in Society "the rich get richer, and the poor get poorer" \cite{merton1968matthew}. The analogy in scientific publishing is the following: In most databases where research is accessed, the papers that have the most citations appear first, and are therefore more likely to be accessed and cited themselves, perpetuating the self-serving cycle. Inevitably, this also leads to papers with lower citation counts being left behind. By taking into account both the number and the provenance of citations, the $i^3$ can limit the occurrence of this phenomenon. Indeed, some publications with lower citation counts may still fare up at the top if the citations are found in high-impact journals or in JCR categories that have a low number of journal titles. 
\\

The strengths of the $i^3$ are also multiple. First, the $i^3$ is the only standardized article-level metric that uses and distinguishes citation provenance. Secondly, unlike \textit{altmetrics}, the $i^3$ focuses solely on scientific references, giving credence to the notion that the impact it seeks to evoke comes from rigorous sources. Thirdly, the $i^3$ is author-blind and independent, and compared to \textit{altmetrics}, where authors can tweet and access their own work, the $i^3$ is less amenable to self-intervention by the authors. But perhaps the biggest strength comes from the simplicity of its nature: a simple number between 0 and 1 that is easily used to compare individual publications. 
\\

On the other hand, the $i^3$ does have a number of limitations. First, citations can come from journals in different disciplines with highly variable impact factors, which may introduce bias. Similarly, not all press is good press: some citations are actually critiques and may actually imply that the article in question evokes the opposite of 'quality'. Furthermore, while the $i^3$ is based on the assumption that citations in prestigious journals with greater readership implies higher potential for impact, there is no guarantee that readers will actually engage with bibliographic material. In addition, the $i^3$ does not distinguish between years of publication, self-citations or order of authorship. And finally, the $i^3$ may not be a reflection of an article's worth, but rather of the topic at hand, the accessibility of the publishing journal, or the popularity of the journal where it is published.
\
\subsection{How to use the $i^3$}
Imagine that, as a reader of Science, you are interested in reading about a particular subject, or answering a specific question on a given topic. How do you go about finding the right literature? Most Scientists will access official, well-known databases, where the subject's literature is likely to be found. Once in the database, one may limit the vast number of results by performing an advanced search, limiting findings by language, year of publication, or authorship, amongst others. Given the vast literature that is available, dozens, perhaps hundreds or thousands of results will be available. The task of going through all of that literature is daunting and certainly time-consuming. Having an $i^3$ value associated with each publication would provide a rapid and simple way to compare the impact of individual publications, serving as an additional way to restrict findings and increase the likelihood of finding the right literature to answer their query. By simply comparing two or more $i^3$ values, simple numbers between 0 and 1, and ranking publications from the highest $i^3$ to the lowest, the reader can guide his search, by counting on the validation of their peers who, through citations and perpetuation of findings, have endorsed the scholarly work in question. 

\section{Conclusion}

The $i^3$ is a novel article-level metric that can be used to assess and compare a standardized measure of the impact of individual scholarly publications. In addition to its main objective, the adoption of this new metric may have implications in hiring practices, distribution of funds and grants, and ranking scientists' portfolios at large.

\newpage

\section{Addendum}
\subsection{Balayla ($\beta$) Coefficients in sample JCR categories}
\begin{center}

\begin{table}[h]
\centering
\
\
\begin{tabular}{lcccl}
\hline
 & \textbf{JCR Category}     & \textbf{Number of titles ($\phi$)} & \textbf{Balayla Coefficient ($\beta$)} &  \\ \hline
 & Astronomy \& Astrophysics & 61                             & 0.001739398                  &  \\
 & Plant Sciences            & 209                            & 0.000507671                  &  \\
 & Immunology                & 150                            & 0.000707355                  &  \\
 & Neurosciences             & 256                            & 0.000419381                  &  \\
 & Pharmacology \& Pharmacy  & 253                            & 0.000419381                  &  \\ \hline
 & ...                       & ...                            & ...                          & 
\end{tabular}
\end{table}
\end{center}
\subsection{Plot of varied $\beta$ coefficients in independent $i^3$ functions}
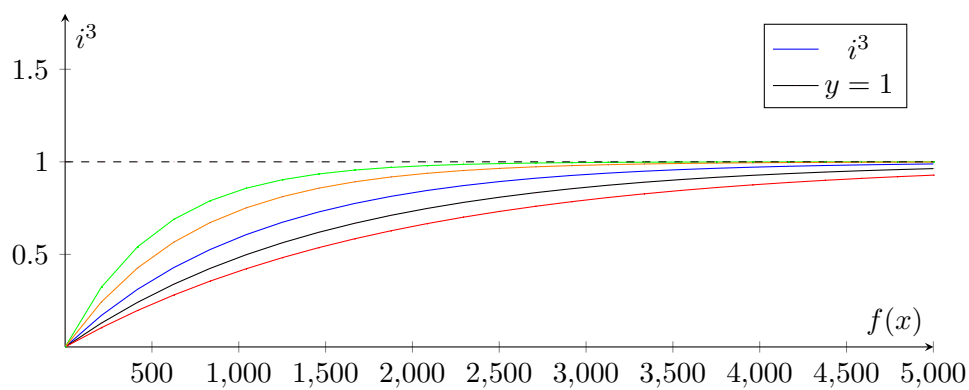
\begin{center}

\begin{tikzpicture}
 
	\begin{axis}[
    axis lines = center,
    xlabel = $f(x)$,
	ylabel = {$i^3$},    
     ymin=0, ymax=1.8,
    legend pos = north east,
     ymajorgrids=false,
    grid style=dashed,
    width=13cm,
    height=6cm,
     ]
	\addplot [	
	domain= 0:5000,
	color= blue,
	]
	{1-2.1778^(-0.00115*x)};
\addplot+ [
domain= 0:5000,	
color = black,
mark size = 0pt
 ]
	{1-2.1778^(-0.00085*x)};
\addplot+ [
domain= 0:5000,	
color = red,
mark size = 0pt
 ]
	{1-2.1778^(-0.00067643*x)};	
	
	\addplot+ [
domain= 0:5000,	
color = green,
mark size = 0pt
 ]
	{1-2.1778^(-0.0024*x)};
	\addplot+ [
domain= 0:5000,	
color = orange,
mark size = 0pt
 ]
	{1-2.1778^(-0.00171543*x)};
\addplot+ [
domain= 0:5000,
dashed,	
color = black,
mark size = 0pt
 ] 
	{1};	
	\addlegendentry{$i^3$}
	\addlegendentry{$y=1$}
	\end{axis} 

\end{tikzpicture}
\end{center}

\begin{center}
Each $i^3$ curve rises at different velocities. This is the nature of the \textit{Balayla} coefficient. Nonetheless, note how independent of the $\beta$ coefficient, all $i^3$ formulations have an asymptote at y=1.
\end{center}

\newpage

\bibliographystyle{apalike}
\bibliography{references}

\end{document}